\documentclass[aps,prd,tightenlines,nofootinbib,amsfonts,amssymb,amsmath,11pt]{revtex4-1}
\usepackage{graphicx}  
\usepackage{color}
\usepackage{hyperref}

\begin{document}

\newcommand{\etal}{{\it et al.}}
\newcommand{\bx}{{\bf x}}
\newcommand{\bn}{{\bf n}}
\newcommand{\bk}{{\bf k}}
\newcommand{\dd}{{\rm d}}
\newcommand{\dslash}{D\!\!\!\!/}
\def\ga{\mathrel{\raise.3ex\hbox{$>$\kern-.75em\lower1ex\hbox{$\sim$}}}}
\def\la{\mathrel{\raise.3ex\hbox{$<$\kern-.75em\lower1ex\hbox{$\sim$}}}}
\def\beq{\begin{equation}}
\def\eeq{\end{equation}}
\def\be{\begin{equation}} 
\def\ee{\end{equation}}
\def\bea{\begin{eqnarray}}
\def\eea{\end{eqnarray}}

\newcommand{\dif}{\mathrm{d}}
\newcommand{\me}{\mathrm{e}}
\newcommand{\mi}{\mathrm{i}}
\newcommand{\bs}[1]{\boldsymbol{#1}}
\newcommand{\de}{\partial}
\newcommand{\mpl}{M_{\rm Pl}}

\def\C{{\cal C}}
\def\cleeu{{\tilde \C}_l^{EE}}
\def\clteu{{\tilde \C}_l^{TE}}
\def\clttu{{\tilde \C}_l^{TT}}
\def\a{\alpha}
\def\im{\eta^{-1}}

\def\tr{{\rm tr}}


\vskip-2cm
\title{Generalised tensor fluctuations and inflation}
\author{Dario Cannone$^{(a,b,c)}$, Gianmassimo Tasinato$^{(c,d)}$, David Wands$^{(c)}$}

\affiliation{
$^{(a)}$ Dipartimento di Fisica e Astronomia G. Galilei \\
Universit\`a degli Studi di Padova, I-35131 Padova, Italy.
\\
\hskip0.7cm
\\
$^{(b)}$
INFN, Sezione di Padova, I-35131 Padova, Italy.
\\
\hskip0.7cm
\\
$^{(c)}$
Institute of Cosmology \&  Gravitation, University of Portsmouth, Dennis Sciama Building, Burnaby Road, Portsmouth, PO1 3FX, United Kingdom
\\
\hskip0.7cm
\\
$^{(d)}$
Department of Physics, Swansea University, Swansea, SA2 8PP, UK
\hskip0.7cm
}

\begin{abstract}
Using an effective field theory approach to inflation,  we examine novel properties of the spectrum of inflationary tensor fluctuations, that arise 
 when breaking some of the symmetries or requirements  usually imposed on the dynamics of perturbations. 
During single-clock inflation, time-reparameterization invariance 
is broken by a time-dependent cosmological background.
In order to explore more general scenarios, we consider the possibility
that spatial diffeomorphism invariance is also broken by effective mass terms
or by derivative operators for the metric fluctuations in the Lagrangian. 
We investigate the cosmological consequences of the       
           breaking of spatial diffeomorphisms, focussing on operators that  affect the power spectrum of fluctuations. 
We identify the operators for tensor fluctuations that can  provide a blue spectrum without violating the null energy condition, 
and operators for scalar fluctuations that lead to non-conservation of the comoving curvature perturbation on superhorizon scales 
even in single-clock inflation.
In the last part of our work, we also examine the consequences of operators containing
           more than two spatial derivatives, discussing how they
           affect the sound speed of tensor fluctuations, and  
           showing that they can mimic some  of the interesting effects of 
   symmetry breaking operators, even in scenarios that preserve spatial diffeomorphism invariance. 
\end{abstract}
\maketitle

\section{Introduction}

The recent  results from the BICEP2 collaboration \cite{Ade:2014xna} suggest that 
 CMB polarization measurements are  reaching sufficient sensitivity to start  detecting primordial 
 B-modes, if foregrounds can be understood and the 
  gravity wave
   amplitude is sufficiently large. In this optimistic situation, recent theoretical studies \cite{Caligiuri:2014sla,Dodelson:2014exa,Boyle:2014kba} suggest  that if a sufficient delensing 
   of the B-mode signal 
   can be performed,
  then both the tensor-to-scalar ratio $r$, and the tilt of the tensor spectrum $n_T$ might  be measured with an
  accuracy 
  sufficient   to test the consistency relation
  \be n_T=-r/8\,,\label{sfpred}\ee  
  that holds 
   for
  standard single clock inflation  in Einstein gravity.
     
 This motivates a general theoretical investigation of  possible mechanisms for producing
  primordial tensor fluctuations 
 during inflation,  including  scenarios that are more general  than the ones studied so far.    
 A generic prediction of standard single-field, slow-roll inflation is the production of a nearly scale invariant spectrum of tensor
modes with an amplitude
 proportional to the Hubble parameter during inflation, a ratio $r<1$ between the tensor and scalar power spectra,  and a tilt $n_T<0$ of the tensor spectrum related to $r$ by eq \eqref{sfpred}: see e.g. \cite{Lidsey:1995np} for a review.
  The single clock consistency relation \eqref{sfpred} can be violated in multiple field models (see \cite{Bassett:2005xm}
 for a review); however,  in inflationary scenarios 
 based on a slow-roll expansion, that do not violate the Null Energy Condition, $n_T$ is generically negative.  On
 the other hand, various specific examples have been proposed in the literature that are able to   obtain a
   positive $n_T$   in a controllable way. One can 
   include to eq \eqref{sfpred} contributions that are  higher order slow-roll   \cite{Gong:2014qga},  or
  violate  the Null Energy Condition in Galileon or Hordenski constructions \cite{Kobayashi:2010cm}. Alternatively, 
  one can consider   particle production during inflation
\cite{Mukohyama:2014gba}, or investigate specific non-standard scenarios as 
  solid/elastic inflation \cite{Endlich:2012pz,Gruzinov:2004ty}.



 In this work, we take a more general perspective to the problem of characterizing tensor fluctuations.
  By implementing
 an effective field theory approach to  inflation, 
 we examine novel properties of the spectrum of inflationary tensor fluctuations, that arise 
 when breaking some of the symmetries or requirements  usually imposed on the dynamics of perturbations. 
During single-clock
 inflation, the time-diffeomorphism invariance
  is normally broken by the time dependent cosmological background configuration: the construction of the most general
  theory for fluctuations that preserves spatial diffeomorphisms, but  breaks the time reparametrization 
  invariance, leads to 
   the effective theory 
    of single field inflation initiated in  \cite{Cheung:2007st}, and developed 
     by many groups over the past few years (see \cite{Tsujikawa:2014mba} for a recent review on this topic).

    On the other hand, 
    it might very well be possible that
    during inflation also the {\it spatial} diffeomorphism invariance 
   is  broken in the lagrangian for fluctuations.
      This possibility has not been much explored 
    in the literature, apart from interesting specific set-ups as solid inflation
    \cite{Endlich:2012pz}.  
     Alternatively,   
    operators with more than two spatial derivatives  acting on the tensor perturbations --  preserving or not spatial diffeomorphism invariance -- could become important in situations where the leading order Einstein-Hilbert contributions to the tensor sector can be neglected, and provide interesting contributions to inflationary
    observables.    
    
  In this article, we   explore these possibilities using an effective field theory approach.
     We consider the dynamics of   metric fluctuations for single clock inflation in a unitary gauge
      in which the  clock perturbations
     are set to zero, and for simplicity we   concentrate on   operators
      that are at most quadratic in fluctuations,  since our   main aim 
   is  to try to understand how they can affect  observables such as $r$ and $n_T$, that are 
   directly associated with the tensor power spectrum.
      
    In the
   first part of the work, we   study    contributions
   to the effective lagrangian for perturbations  that break the  spatial diffeomorphism invariance by effective mass terms, 
       or by derivative operators for  the metric fluctuations. 
        In order not  to induce spatial anisotropies, we  limit our attention to contributions that do not break
         the Euclidean symmetry in the spatial sections, 
 corresponding to an            
         $SO(3)$ rotational invariance.  
       We study   the conditions
         one has to satisfy to avoid ghost instabilities and   to have well-behaved fluctuations; moreover  we examine some
         cosmological consequences of our findings. 

          In the second part of this work,  we study operators quadratic  on the metric  fluctuations that contain up to
          four spatial derivatives (but no more than two time derivatives), 
          that can  preserve spatial diffeomorphism invariance, and that can have interesting effects in regimes where 
          they provide the dominant contribution to the  tensor dynamics. We show that a non-trivial tensor sound speed can be
          generated, and the formula for the 
            tilt of the tensor spectrum  receives  new  contributions that depend  on
             the coefficients of these higher derivative operators. In particular, we discuss
             a special case in which  
                  such operators can mimic the effect 
          of a mass term in the tensor sector. 


 We do not wish to systematically investigate 
  {\it all} possible operators  with the properties we are  interested in, but to study  representative and promising 
  examples 
   that can  be of some use to connect 
    inflationary model building   with observations, especially when  focussing on the tensor sector. 
   On the other hand, the tools that we develop can be further applied and generalized to study more general
   situations, for example in set-ups with broken  isotropy in the effective action for fluctuations.   
   Since we implement an effective field theory approach to the study of perturbations from inflation, we do not attempt to find actual theories or models whose 
    cosmological 
   fluctuations have the properties we investigate, although  we will  also comment on possible
     realizations  for the operators we  study.  
 We limit our attention to operators that are quadratic in fluctuations. Given the fact that we break some of
  the symmetries such as spatial diffeomorphism invariance, 
many operators cubic or higher in fluctuations exist;  this considerably complicates a systematic analysis of their effects, that
   we leave for  future work.

\section{Breaking spatial diffeomorphism invariance}

In this section we investigate  an effective field theory for cosmological perturbations around quasi-de Sitter space,  with broken spatial and time diffeomorphism invariance.

We take a conformal (FRW) ansatz for the background metric,
\be\label{backcon}
d s^2\,=\,\bar{g}_{\mu\nu}\,d x^\mu d x^\nu\,=\,a^2(\eta)\,\left( -\eta_{\mu \nu}\,d x^\mu d x^\nu \right)
\ee
 with $a^2(\eta)$ the conformal scale factor and $a(\eta)\,=\,1/(-H \eta)$ for de Sitter space.  
 We denote the metric fluctuations by
  $h_{\mu\nu}=g_{\mu\nu}-\bar{g}_{\mu\nu}$ .
  
  The  time-reparameterization invariance for fluctuations is 
      broken by the time dependence of the homogeneous background. 
  In addition, we would like to study the effects of breaking spatial diffeomorphism invariance.  
  The breaking of diffeomorphism invariance
   in the spatial sections is most easily achieved by mass terms, although derivative operators
   involving  metric pertubations  are also able
   to do so.
 
    First we consider the effects  of mass terms, before
    including  diffeomorphism-breaking derivative operators in the next subsections. 
    These operators corresponding to mass terms do not necessarily originate by a theory of massive gravity holding
    during inflation; they simply correspond to the most general way to express quadratic  
     non-derivative operators in the fluctuations that break diffeomorphism invariance. 
     
    
      We consider the 
    Einstein-Hilbert action expanded to second order, and add 
    generic operators with no derivatives, that are  quadratic in the metric fluctuations $h_{\mu \nu}$ 
\begin{equation}
 \begin{array}{lcl}
  S & = & \displaystyle\int \dif^4x\sqrt{-g}\mpl^2\Bigg[R-2\Lambda -2\,c\, g^{00}\Bigg]+ \\
    &   & \displaystyle\frac{1}{4}\mpl^2\int\dif^4x\sqrt{-g}\Bigg[m_0^2 \, h_{00}^2+2 \, m_1^2 \, h_{0i}^2-m_2^2 \, h_{ij}^2+m_3^2 \, h_{ii}^2-2m_4^2 \, h_{00} \, h_{ii}\Bigg]\,.\label{act1}
 \end{array}
\end{equation}
The terms in the first line are the ones that will give the homogeneous and isotropic background which we assume for inflation.
They give a non-zero stress-energy tensor at background level,
\begin{equation}
 T^{(0)}_{\mu\nu}=-\frac{2}{\sqrt{-g}}\frac{\delta S}{\delta g^{\mu\nu}}\Bigg|_{background}
\end{equation}
and, using Friedmann equations, the parameters $c$ and $\Lambda$ can be expressed as functions of the Hubble parameter $H$ and its time derivative $\dot{H}$ (that defines the slow-roll parameter $\epsilon\,=\,-\dot{H}/H^2$).

The quadratic terms in the second line of Eq.~(\ref{act1}) break diffeomorphism invariance,  yet they preserve a spatial  $SO(3)$ invariance in order not to break spatial isotropy.
 The term proportional to $m_0^2$ breaks time reparameterization invariance, and is present also in the 
 quadratic Lagrangian of \cite{Cheung:2007st}: the remaining terms in the second line of 
 Eq.~\eqref{act1}, instead, are absent in \cite{Cheung:2007st}, and break spatial diffeomorphism invariance. 
 They have the same structure of the Lorentz
violating  mass terms of \cite{Dubovsky:2004sg}, this time applied to the case of an expanding (quasi)-de Sitter
universe. They were dubbed `Lorentz violating' in \cite{Rubakov:2004eb,Dubovsky:2004sg} since in the flat limit ($H\to0$) they do break 4d Lorentz symmetry  $SO(1,3)$ down to spatial rotational
symmetry, $SO(3)$ \footnote{
For certain choices of the parameters, these mass terms (although breaking diffeomorphism invariance)
can recover 4d Lorentz invariance in the flat limit $H\to0$. 
The parameter choice one has to make is
 \begin{equation}
 m_0^2=\alpha+\beta\,,\qquad m_1^2=m_2^2=-\alpha\;,\qquad m_3^2=m_4^2=\beta \; .
\end{equation}
and  the Fierz-Pauli theory corresponds to  $\alpha+\beta=0$.  These arguments are reviewed in \cite{Rubakov:2008nh}
.}.  Since the choice of operators we consider
preserves isotropy  at each point in space, they also preserve homogeneity in space.
In the limit $m_i\to0$ with $i\neq0$, spatial diffeomorphisms are restored and, up to second order in perturbations,
we recover the standard effective field theory of inflation \cite{Cheung:2007st} without  extrinsic curvature terms,
where only time diffeomorphisms are broken by powers of $h_{00}$.



  We can consider the  `mass terms' in the second line of Eq.~(\ref{act1}) as arising from couplings between
the  metric and fields acquiring a nontrivial time-dependent profile during inflation.
 We assume that their coefficients (as well as the ones that we will meet in the following)
 are effectively constant
in space and time during inflation,  while these coefficients  go to zero after inflation,  
and hence are not constrained by present day observational limits.  
The constancy in space is not a strict requirement since effects of  gradient terms are usually negligible
at large scales during inflation. A (small) time dependence for these operators would instead be
expected, proportional to slow-roll
parameters quantifying the departure from an exact de Sitter phase during inflation: for simplicity we will neglect it.  

We will not consider interactions in this paper, but we will limit our attention to terms quadratic
 in perturbations.  Nevertheless, for the class of mass terms contained in action~(\ref{act1}),
  general considerations show 
    that the maximal cut-off is of order $\Lambda_c \simeq \sqrt{m\,M_{Pl}}$ \cite{ArkaniHamed:2002sp}, 
     assuming that all the non-vanishing
   mass parameters are of the same magnitude $m$.  In order to have a reliable theory, we must ensure that $\Lambda_c\ge H$, where $H$ is the Hubble scale during inflation,  so that
\be\label{cocond}
\frac{m}{H}\ge \frac{H}{M_{Pl}} \,.
\ee
Hence for inflation happening at high energy scales, 
  the mass of the graviton must be quite large during the inflationary process  (although
it can be well below the Hubble scale). After inflation ends, we assume that the effective 
graviton mass becomes negligible,  as we mentioned above. 


Let us stress that 
 in the spirit of our   effective approach to cosmological fluctuations, only based on symmetry arguments,  it is {\it 
 not necessary} to specify the nature
of the model (the `UV completion') that leads to the fluctuation Lagrangian we are examining. 
Our theory appears as a version of (Lorentz violating) 
 massive gravity because 
   we are selecting a specific gauge -- the unitary gauge -- in which fluctuations of the field(s) driving
   inflation are set to zero: the  dynamics of  perturbations is entirely described by the sector of  metric fluctuations. Depending on the set-up under consideration, other gauges could  be chosen though, in which the graviton is massless,  and other sectors play the role in determining the dynamics of     fluctuations during inflation.
 
The UV completion of our scenario 
  might be some
 specific version of massive gravity coupled to an inflaton field (for reviews of massive gravity, see e.g. \cite{Rubakov:2008nh,deRham:2014zqa}), or some model of inflation
  making use of vectors (see \cite{Maleknejad:2012fw} for a review), or sets of scalars obeying specific symmetries. For example,  solid inflation \cite{Endlich:2012pz} is a set-up with broken spatial diffeomorphisms (but preserving time-reparameterization); 
   the dynamics of its fluctuations 
      might be considered
  as a subclass of our  general discussion.       
    


\subsection{Tensor-vector-scalar decomposition}

It is helpful to rewrite the action (\ref{act1}) in terms of tensor, vector and scalar perturbations on spatial hypersurfaces, which evolve independently at linear order:
\begin{equation}
 \begin{array}{lclllll}
  h_{00} & = & \psi \,,& & & \\
  h_{0i} & = & u_i+\de_iv \,, & \hskip0.2cm {\rm with} \hskip0.2cm& \de_iu_i & = & 0 \,,\label{deco}\\
  h_{ij} & = & \chi_{ij}+\de_{(i}s_{j)}+\de_i\de_j\sigma+\delta_{ij}\tau\,, &\hskip0.2cm {\rm with}\hskip0.2cm& \de_is_i & = &\de_j\chi_{ij}=\delta_{ij}\chi_{ij}=0\,. 
 \end{array} 
\end{equation}
Under a diffeomorphism,
$\eta\to \eta+\xi_0$, $x^i\to x^i+\xi^i$, 
these perturbations transform as
\bea
   \chi_{ij} && \to \chi_{ij} 
\nonumber
\\
    u_i && \to u_i + \partial_0 \xi^T_i
\nonumber
\\
    s_i && \to s_i + \xi_i^T
\nonumber \\
   \psi && \to \psi + 2 \partial_0 \xi_0+2aH\xi_0
\nonumber
\\
   v && \to v+ \partial_0 \xi^L + \xi_0 \nonumber
\\
   \sigma && \to \sigma + 2\xi^L
\nonumber
\\
   \tau && \to \tau  +2aH\xi_0 \label{diffeotr}
\eea
where $\xi_i=\xi^T_i+\de_i\xi^L$.
 Expanding (\ref{act1}) up to second order in these fluctuations, we find the following tensor-vector-scalar actions
 including the mass terms:
 \begin{itemize}
 \item[-] Tensor action
 \begin{equation}\label{S_tens}
                 S_{m}^{(T)}=\frac{1}{4}\mpl^2\int\dif^4xa^2\Bigg[-\eta^{\mu\nu}\de_\mu\chi_{ij}\de_\nu\chi_{ij}-a^2m_2^2\chi_{ij}^2\Bigg] \; ,
                \end{equation}
 \item[-] Vector action
  \begin{equation}\label{S_vec}
                 S^{(V)}_{m}=\frac{1}{2}\mpl^2\int\dif^4xa^2\Bigg[-(u_i-s'_i)\nabla^2(u_i-s'_i)+a^2(m_1^2u_i^2+m_2^2s_i\nabla^2s_i)\Bigg] \; ,
                \end{equation}
 \item[-] Scalar action
  \begin{equation}\label{S_scal}
                 \begin{array}{lcl}
                 S^{(S)}_m&=\displaystyle\frac{1}{4}\mpl^2\int\dif^4x\,a^2&\!\!\!\!\!\Bigg\{-6(\tau'+aH\psi)^2+2(2\psi-\tau)\nabla^2\tau+4(\tau'+aH\psi)\nabla^2(2v-\sigma') \\
                        & & +a^2\Big[(m_0^2+2\epsilon H^2)\psi^2-2m_1^2v\nabla^2v-m_2^2(\sigma\nabla^4\sigma+2\tau\nabla^2\sigma+3\tau^2) \\
                        & & +m_3^2(\nabla^2\sigma+3\tau)^2-2m_4^2\psi(\nabla\sigma+3\tau)\Big]\Bigg\}
                 \end{array}
                \end{equation}
 \end{itemize}
 
 Since  diffeomorphisms are broken,
 one would expect to find   six propagating  degrees of freedom, and  one of these should
 generically  be a ghost. 
  Nevertheless, it has been shown that in a FRW background the theory can be ghost-free, and   potential instabilities
  avoided,  if the masses $m_i$ satisfy certain conditions \cite{Blas:2009my}.  In the next subsections,
  we will generalize this analysis including also the effect of 
   a selection of 
  derivative operators that break diffeomorphism
  invariance, studying each sector of the theory and also discussing possible phenomenological consequences. To the operators considered so far we will add new quadratic operators that contain at most two space-time
  derivatives in $h_{\mu\nu}$.  They potentially break spatial diffeomorphism invariance, although they preserve Euclidean invariance
  in the spatial sections. See Appendix \ref{appA} for a list of such operators. To conclude  this section, let
  us point out that
   our analysis 
   includes operators with higher spatial derivatives acting on the fields obtained
   after the  tensor-vector-scalar decomposition of $h_{\mu\nu}$
    (see for example
  the $m_2^2$ coefficient in eq. (\ref{S_scal})) that have been removed by a parameter choice   in \cite{Gleyzes:2013ooa}.  See however \cite{Kase:2014cwa} for a recent analysis including operators   that are higher
  order in spatial derivatives.

\subsection{Tensor Fluctuations}

Let us start by discussing the tensor fluctuations, since this is  the sector we are most interested in. 
We see from the action $S_m^{(T)}$ in Eq.~\eqref{S_tens}  that tensors acquire a  mass  only in the case $m_2^2\neq0$ and no instabilities
arise 
 if $m_2^2\geq0$. Hence only the operator proportional to $m_2^2$ in Eq.~\eqref{act1} influences the tensor spectrum
 by giving an effective mass to the tensors.
  On the other hand, we can
  add to the mass term additional operators
   that contain up to two space-time derivatives and preserve  isotropy: 
   they can change speed of sound for tensor perturbations in
    eq. \eqref{S_tens}. In particular, the only  allowed operators    that can contribute to the tensor sound
     speed are the ones  in 
eqs. \eqref{b1}, \eqref{d1} in appendix \ref{appA}. 

We may add to the action \eqref{S_tens} two derivative operators~\footnote{Notice that also
  a parity violating, one derivative operator could be included, $\epsilon^{ijk} \left(\partial_i\,h_{jm}\right)\,
  h_{km}$, with $\epsilon^{ijk}$ the  totally antisymmetric operator in  three spatial dimensions.  On the other
  hand, in this work we concentrate on operators that preserve parity, so we do not consider its effects. We thank  Azadeh Maleknejad
  for discussions on this point.}, with dimensionless
 coefficients $b_1$ and $d_1$:
\begin{equation} \label{derct}
S^{(T)}_{d}\,\equiv\,
 \frac{1}{4}\mpl^2\left[b_1(\de_0 h_{ij})^2+d_1(\de_i h_{jk})^2\right] \;.
\end{equation}
It is important to notice that these two derivative operators do not necessarily originate from contributions that  break the
3-dimensional diffeomorphism invariance {\it per se}. In particular these terms can arise from the diffeomorphism invariant combination
$b_1\delta K_{ij}\delta K^{ij}-d_1\!^{(3)} R$, where $\delta K_{ij}\delta K^{ij}$ is the perturbed extrinsic curvature and $^{(3)}\delta R$
is the three-dimensional Ricci scalar \cite{Cheung:2007st, Noumi:2014zqa}.
   These specific combinations, 
  on the other hand, contain specific additional vector and scalar contributions that have to be taken into account.
  We will consider them in the next subsections, but for the moment we do not need to 
  restrict to any special case; we can focus on (\ref{derct}) regardless of its origin. 
  
The complete action for tensor fluctuations becomes
\begin{equation}\label{tens_derct}
 S^{(T)}\,=\,
 S^{(T)}_{m}+ 
 S^{(T)}_{d}
 \,=\,\frac{1}{4}\mpl^2\int\dif^4xa^2\Bigg\{(1+b_1)\Big[(\dot\chi_{ij})^2 - c_T^2(\de_i\chi_{jk})^2\Big]
         -a^2m_2^2\chi_{ij}^2\Bigg\} \; ,
\end{equation}
where the speed of sound for tensors is
\begin{equation}
 c_T^2=\frac{1+d_1}{1+b_1} \;.
\end{equation}
In this case, in order to avoid ghosts one should also require $b_1>-1$, $d_1\geq-1$; moreover
we could also demand $d_1\leq b_1$ not to have superluminal propagation.


Taking the action \eqref{tens_derct}, it is easy to derive the expression for the tensor power spectrum, 
quantizing the tensor fluctuations starting from the usual Bunch-Davies vacuum. 
 Upon canonical
 normalization and neglecting for simplicity time dependencies of $c_T$ and $m_2$, the equation of motion 
  for tensors 
 has the usual
 Mukhanov--Sasaki form. It can be solved to give the following  expression for the power spectrum and its
 scale dependence:
 \begin{equation} \label{ptsca}
  \mathcal{P}_T=\frac{2H^2}{\pi^2\mpl^2c_T}\left(\frac{k}{k_*}\right)^{n_T} \;,\qquad
  n_T=-2\epsilon+\frac{2}{3}\frac{m_2^2}{(1+b_1)^2H^2}\left(1+\frac{4}{3}\epsilon\right) \;,
 \end{equation}
at leading order in slow-roll and in an expansion in $m_2/H \ll 1$.  Notice that the mass term can render 
the tensor spectrum blue if $m_2/H$ is sufficiently large and positive so that the second term in $n_T$ wins out over the negative
contribution from the first term.  
 In this sense, a blue spectrum
 for tensors can be obtained without violating the Null Energy Condition or exploiting the  
 time-dependence of parameters: it is the effect of  the mass term proportional to $m_2^2$ and is
 not depending on the sign of $\dot{H}$.   
 
 It would be interesting to explore 
 whether if we choose different initial conditions that do not preserve isotropy, then the operators that we consider would lead to
 an anisotropic signal during inflation, as happens in the particular set-up of solid inflation
 \cite{Bartolo:2013msa,Bartolo:2014xfa}, both in the tensor and in the scalar and vector sectors. This will be
 the subject of future work \cite{preparation}.

 
 The amplitude of the tensor power spectrum is enhanced by the inverse of the sound speed $c_T$.
 On the other hand,
  it has been recently shown in \cite{Creminelli:2014wna} that, 
  when focussing on operators containing at most two derivatives -- as we do in this section --
   there exists a disformal
redefinition of the metric which converts the theory with a speed of sound $c_T\neq1$ into a theory (in
the Einstein frame) 
 with unit speed
of sound. Thus,  in the Einstein frame, during inflation
 the sound speed is equal to one. Hence -- neglecting the scale dependence of ${\cal P}_T$ --  the amplitude of the tensor power spectrum is directly linked to the scale of inflation.
  Notice that in our scenario  
  we do have an additional source of scale-dependence though, associated with the mass term $m_2$
  that breaks the spatial diffeomorphism invariance. The disformal
  transformation of \cite{Creminelli:2014wna} does not involve spatial coordinates hence does not modify our
  predictions for the scale dependence of the tensor spectrum, whose
  sign is still controlled by $m_2^2/H^2$ versus $\epsilon$. 
  
  It has been discussed in 
  Ref.~\cite{Creminelli:2014wna} that terms involving higher derivatives can actually change 
   the situation and induce a non-trivial sound speed. While in  \cite{Creminelli:2014wna} three-derivative
   terms were included, we will extend this possibility  and study  healthy four derivative terms (with at most
   two time derivatives) in Section \ref{4derivatives}.

\subsection{Vector Fluctuations}

We now discuss the propagation of vector fluctuations in our set-up. In this and in the next subsection  (where we will discuss the dynamics of  scalars) we do not pretend to be exhaustive in our analysis, but only to investigate simple and interesting cases among the many possibilities allowed within our large  parameter space. In particular, aiming for simplicity, our purpose is to reduce as much as we can the number of propagating degrees of freedom in our scenario, and choose parameters which can eliminate the vector degrees of freedom. We will study the general case in   \cite{preparation}.   
    
 
In principle we have two vector degrees of freedom, $u_i$ and $s_i$, from the decomposition in eq (\ref{deco}). 
Examining the action (\ref{S_vec}) for vector perturbations including mass terms, 
 and in absence of additional derivative operators,  it  is straightforward
  to show that the field $u_i$ is not dynamical, since we obtain
\begin{equation}
 \nabla^2(u_i-s'_i)-a^2m_1^2u_i=0 \,.
\end{equation}
Hence $u_i$ can be integrated out to give the effective action
\begin{equation}\label{vem2}
 S^{(V)}_m=\frac{1}{2}\mpl^2\int\dif^4x\,a^4\Bigg[m_1^2s_i'\frac{\nabla^2}{\nabla^2-a^2m_1^2}s_i'+m_2^2s_i\nabla^2s_i\Bigg] \; . 
\end{equation}
The action is free of instabilities for
 $m_1^2\geq0$ 
and $m_2^2\geq0$.
The case $m_1^2=0$ is particularly interesting as there are no propagating vector modes, since the 
coefficient of the $s_i$ kinetic term in (\ref{vem2})  vanishes. Hence in order to eliminate vector degrees of freedom, 
we make the choice $m_1=0$. 


On the other hand,
 the situation can drastically change if also other possible derivative contributions are included in the action,
 choosing from the list of allowed operators in Appendix \ref{appA}. 
 There are six possible terms with up to two derivatives that
contribute to the vector sector, that contribute to an effective Lagrangian that we dub ${\cal L}^{(V)}_d$:
\bea
{\cal L}^{(V)}_d&=&
 \frac{1}{4}\mpl^2\,\Big[b_1(\de_0h_{ij})^2+b_2(\de_ih_{0j})^2+b_3(\de_jh_{0i}\de_0h_{ij})
  \nonumber
 \\
 &&
 +d_1(\de_ih_{jk})^2+d_2(\de_ih_{ij})^2\Big]
 \nonumber
 \\
 &&
 +\frac{1}{4}\mpl^3\,\alpha_4\,(h_{ij}\de_ih_{0j})\;,\label{derivvec}
\eea
where $b_i$, $d_i$ and $\alpha_4$ are arbitrary constant coefficients.  Notice that also a single
derivative term is allowed   in the last line of eq (\ref{derivvec}). 

 These derivative contributions in $S^{(V)}_d$ 
  in general 
  switch on a non-trivial dynamics for $s_i$ even if $m_1^2=0$. On the other hand,
  it can be shown (c.f., appendix \ref{appA})
that if one chooses the particular values
\begin{equation}\label{b_combination}
 b_1=\frac{1}{2}b_2=-\frac{1}{4}b_3 \; ,
\end{equation}
 then the structure of the action \eqref{S_vec} would be unaltered and the vector $s_i$, when $m_1^2=0$,  would still be
 non-dynamical. This corresponds  to a combination of the operators forming the spatial diffeomorphism invariant quantity  $(\delta K_{ij})^2$. 
Provided this condition (\ref{b_combination})
is satisfied, adding the  operators proportional to
$d_1$, $d_2$ and $\alpha_4$ in eq. (\ref{derivvec}) does not change the conclusion such that $s_i$ not dynamical.

Hence, the condition  $m_1^2=0$ is appealing since we can still ensure that no vectors propagate. 
As we will see, this condition also gives only one propagating mode in the scalar sector, since extrinsic
curvature terms do not render a second scalar mode dynamical. 
  Of course, other cases can be considered (with propagating vector modes) and our approach 
 will allow us to study them in future \cite{preparation}.
  
Fine-tuning relations on mass parameters, such as $m_1^2=0$ can be motivated and protected
  by residual gauge symmetries \cite{Dubovsky:2004sg}. Indeed, this is the case
for $m_1^2=0$; if we require invariance under time-dependent diffeomorphisms,
\begin{equation}
 x^i\to x^i+\xi^i(t) \;,
\end{equation}
then the operator $h_{0i}$, associated with $m_1^2$, is forbidden in the action.

\subsection{Scalar Fluctuations}
Not surprisingly, the scalar sector is the most tricky to analyze due to the number of fields involved and their
mixings. We separate the discussion in two parts. First we study the case in which only scalar masses are
 included, and no derivative operators are added to eq. (\ref{S_scal}). We show that an important physical
  consequence of our construction is that the curvature perturbation $\zeta$ is generally  not conserved on super-horizon
  scales.  
 We then proceed, including derivative operators in the second part of this section. 

The main aim is to find the conditions required to propagate at most one (healthy) scalar degree of freedom in our system.

\subsubsection{Only  masses are included}

When only scalar masses are switched on, the action we are working with is 
  Eq.~(\ref{S_scal}). This action potentially propagates two degrees of freedom, $\sigma$ and $\tau$. 
It can be shown that even in the case where all the masses are different from zero, the theory has no ghosts nor other instabilities
provided that $m_1^2>0$, $6H^2\geq m_0^2-2\dot{H}>0$ and $\dot{H}<0$ \cite{Blas:2009my}.

Here we focus instead on the case $m_1^2=0$ that, besides having no vectors,
  it
also has only one propagating scalar, as we are going to discuss. From eq. \eqref{S_scal} with $m_1^2=0$ one can obtain the equations of motion for
the auxiliary fields $\psi$ and $v$,
\begin{equation}\label{eom_vpsi}
 \begin{array}{lcl}
  \psi & = & \displaystyle-\frac{\tau'}{\mathcal{H}} \,,\\ & & \\  
  \nabla^2v & = & \displaystyle\frac{a^2}{4\mathcal{H}}\left[(m_0^2-2\dot{H})\tau'-\frac{2}{a^2}\nabla^2\tau+
                  \frac{2\mathcal{H}}{a^2}\nabla^2\sigma'+m_4^2(\nabla^2\sigma+3\tau)\right] \; ,
 \end{array}
\end{equation}
and substitute them back into the action obtaining  (where we write $\mathcal{H}=aH$ and $\dot{H}\,=\,-\epsilon H^2$)
\begin{equation}
 \begin{array}{lcl}
  S=\displaystyle\frac{1}{4}\mpl^2\int\dif^4x & \displaystyle a^2 \Bigg[-2\left(\frac{\tau'}{aH}+\tau\right)\nabla^2\tau+
                                   a^2(m_0^2+2\epsilon H^2)\left(\frac{\tau'}{aH}\right)^2
                                   -a^2m_2^2(\sigma\nabla^4\sigma+2\tau\nabla^2\sigma+3\tau^2) \\
                                &\displaystyle  +m_3^2(\nabla^2\sigma+3\tau)^2+\frac{2m_4^2a^2}{aH}\tau'(\nabla^2\sigma+3\tau)\Bigg] \;.
 \end{array}\label{actws}
\end{equation}
This shows that $\sigma$ is also an  auxiliary field:  
\begin{equation}\label{eom_sigma}
 aH(m_2^2-m_3^2)\nabla^2\sigma=m_4^2\tau'-aH(m_2^2-3m_3^2)\tau \,.
\end{equation}
The action becomes
\begin{equation}\label{S_scal_tau}
 \begin{array}{ll}
  S=\displaystyle\mpl^2\int\dif^4x &\displaystyle\frac{a^2}{H^2}\Bigg[\frac{(m_0^2+2\epsilon H^2)(m_2^2-m_3^2)+m_4^2}{2(m_2^2-m_3^2)}{\tau'}^2
                     +\epsilon H^2\tau\nabla^2\tau \\
                                   & \displaystyle-\frac{m_2^2a^2H^2(m_2^2-3m_3^2+(3+\epsilon)m_4^2)}{m_2^2-m_3^2}\,\tau^2 \Bigg]\\
\,. 
 \end{array}
\end{equation}
After canonical normalization of $\tau$, the action finally is given by 
\begin{equation}
 S=\int \dif^4x a^2\Big[\hat{\tau}'^2+c_s^2(\hat{\tau}\nabla^2\hat{\tau})+a^2M^2\hat{\tau}^2\Big] \;,
\end{equation}
where effective mass and speed of sound are
\begin{eqnarray}
 c_s^2 &=& \frac{2\epsilon H^2(m_3^2-m_2^2)}{m_0^2(m_2^2-m_3^2)+m_4^2} \;, \label{firstcs}\\
 M^2   &=&
  -\frac{2 H^2 m_2^2 \left(m_2^2-3 m_3^2+3 m_4^2\right)}{m_0^2 (m_2^2-m_3^2)+m_4^4} \;, \label{firstM}
\end{eqnarray}
at leading order in slow-roll.

An exhaustive  analysys of all the possibilities for the scalar action is beyond the scope of this work. Other cases besides
the one considered here could be interesting. For example, when $m_1^2=0$ and $m_2^2=m_3^2$, case that 
 is not included in (\ref{S_scal_tau}), 
 it can be shown that no scalar degrees
of freedom propagate \cite{Blas:2009my}. However this is true only if no 
derivative operators 
 for $h_{ij}$ are considered.
When all the other combinations of $h$ and derivatives are considered, they can provide kinetic terms for scalars, changing
the previous conclusions. We will return to this later.

\subsubsection{Non-conservation of ${\cal R}$ and $\zeta$ at super-horizon scales}

Reconsidering   the action \eqref{S_scal_tau}, some interesting points can be made. 
There is only one scalar perturbation, $\tau$, which is related to the comoving curvature perturbation $\mathcal{R}$. In an arbitrary gauge we define
       \begin{equation}\label{comovingR}
        \mathcal{R}=\tau-\frac{\mathcal{H}(\tau'-\mathcal{H}\psi)}{\mathcal{H}'-\mathcal{H}^2} \; .
       \end{equation}
       However in the unitary gauge the equation
       of motion of the auxiliary field $\psi$, eq. \eqref{eom_vpsi}, requires $\tau'=\mathcal{H}\psi$ and we have $\mathcal{R}=\tau$, even when diffeomorphisms are broken
       by the masses. 
       In the limit where all masses go to zero, the scalar
       action \eqref{S_scal_tau} reduces to the standard slow-roll action for $\mathcal{R}$.
      
      Since $\mathcal{R}$ coincides with the (massive) scalar fluctuation $\tau$, $\mathcal R$ (before canonical normalization)   has a non-vanishing
      mass given by 
       \begin{equation}\label{M_onlymass}
        M_\mathcal{R}^2=\frac{m_2^2(m_2^2-3m_3^2+(3+\epsilon)m_4^2)}{m_2^2-m_3^2} \; .
       \end{equation}
       Notice that this mass is present only if $m_2^2\neq0$, exactly as for tensor perturbations.
       A profound implication of this result is that $\mathcal{R}$ is in general \emph{not} constant after horizon exit, as it is in standard
       single-field models of inflation. For $M_\mathcal{R}^2>0$ the solution of the Mukhanov-Sasaki equation for $\mathcal{R}$
       will decay after horizon exit. 
       
       The standard picture of different super-horizon patches of the universe evolving as separate universes with constant $\mathcal{R}$  \cite{Wands:2000dp}
       is not valid anymore.  
        A simple physical interpretation is that, given that diffeomorphism invariance is broken in our set-up, very long wavelength fluctuations
        can no longer be considered as a gauge mode in the zero momentum limit, and there is  actually a  preferred frame (the unperturbed background, 
        $\mathcal{R}=0$) towards
        which the fluctuation dynamics is attracted for $M_\mathcal{R}^2>0$. 
       This is analogous to what happens in  the specific set-up of  solid inflation \cite{Endlich:2012pz}, whose consequences
      can be considered  as a special case of our general discussion. 
 
        Notice that, phenomenologically, in order for the perturbations to remain over-damped on super-horizon scales (not to oscillate and decay rapidly),
       we require $M_\mathcal{R}^2\ll H^2$, which gives a constraint on $M_\mathcal{R}^2$. On the other hand, given that the mass of the tensor depends only
       on $m_2^2$ while the mass of the scalar also on $m_3^2$ and $m_4^2$, there is still enough freedom to have
        a blue tilt for the tensor spectrum and a nearly constant ${\mathcal R}$
 outside the horizon. Actually,        
         making the  particular choice $m_2^2=3m_3^2-(3+\epsilon)m_4^2$ one finds that  ${\mathcal R}$ is 
           massless and conserved outside the horizon.

   In our framework, analogously to solid inflation,  
   the comoving curvature perturbations $\mathcal{R}$ and the curvature perturbations on uniform density slices $\zeta$
       \emph{do not coincide} in the large scale limit, as they do in standard single-field inflation. Indeed, taking the definition
       of the function $\zeta$,
       \begin{equation}
        \zeta=\tau-H\frac{\delta \rho}{\dot\rho} \;,
       \end{equation}
       and computing the density $\rho$ and its perturbation from the energy-momentum tensor, one finds
       at leading order in gradients 
        a contribution that does not
       vanish at large scales:
       \begin{equation}
        \zeta=\tau+\frac{(1-\epsilon)m_4^2}{m_0^2+2\epsilon H^2}\tau+\mathcal{O}(\nabla^2) \neq \mathcal{R}\;.
       \end{equation}
 Also $\zeta$ is not conserved and evolves after horizon exit. Following \cite{Wands:2000dp},
       \begin{equation}\label{zeta_evol}
        \dot{\zeta}=-\frac{H}{\rho+p}\delta p_{\rm nad} +\mathcal{O}(\nabla^2)  \,,
       \end{equation}
       it can be understood that the reason for  this non-conservation is the existence of a non-adiabatic stress induced by
       the presence of the masses. While in the standard case one finds that $\delta p_{\rm nad}$ is proportional only to gradient
       terms, here there is a non-trivial contribution in the perturbed (spatial) energy-momentum tensor even on super-horizon scales, given by
       \begin{equation}
       {\rm Tr}\left[\delta T_{ij}\right]= (m_2^2-3m_3^2)\,{\rm Tr} [h_{ij}] + 3(\epsilon H^2+\frac{1}{2}m_4^2)h_{00} \;.
       \end{equation}
       When diffeomorphisms are preserved, this trace is proportional only to $h_{00}=\psi$, which can then be substituted using
       the constraints \eqref{eom_vpsi} to see that indeed only gradients remain. When diffeomorphisms are broken by the masses,
       the use of the equation of motion \eqref{eom_vpsi} and \eqref{eom_sigma} does not allow us to get rid of all the terms
       and we are left with
       \begin{equation} \label{deltap_nad}
       {\rm Tr}\left[\delta T_{ij}\right]= m_2^2 f(m_i) \tau + \mathcal{O}(\nabla^2) \;.
       \end{equation}
       where $f(m_i)$ is a (complicated) function of all the mass parameters. This term will not vanish on large scales, making $\zeta$
       evolve also after the horizon exit.
  The cause of the non conservation of $\zeta$ and $\mathcal{R}$ has to be understood in 
  terms of the contribution  
       $m_2^2$. Indeed if $m_2^2=0$ curvature perturbations are constant beyond the horizon. The operator proportional to $m_2^2$
       is the only one that gives a non-trivial off-diagonal contribution to the energy-momentum tensor,
       \begin{equation}
        T_{ij}\sim m_2^2 h_{ij} \;,
       \end{equation}
       and hence an anisotropic stress, that is sourced by the very same operator that gives an effective
       mass to the graviton (although
       we will see next that diffeomorphism breaking derivative operators can also play a role). This is coherent and very similar  with what was found in \cite{Endlich:2012pz}, where it is shown that
       a non-vanishing anisotropic stress with certain characteristic on large scale violate some technical assumptions of Weinberg's theorem on the
       conservation of curvature perturbations \cite{Weinberg:2003sw}.

%


\subsubsection{Adding derivative  operators}

Let us now add derivative operators. We by adding  the combination $( \delta K_{ij})^2$ [corresponding to the first line of eq \eqref{derivvec} 
  with the condition
   \eqref{b_combination} for the operators $(\de_0h_{ij})^2$,
$(\de_ih_{0j})^2$ and $(\de_jh_{0i}\de_0h_{ij})$], that as we have seen has the nice feature of 
 avoiding the propagation of vectors. We then 
subtract $(\delta K_{ii})^2$ [including the operators
$(\de_0h_{ii})^2$, $(\de_ih_{0i})^2$ and $(\de_ih_{0i}\de_0h_{jj})$)] in order to avoid the propagation
of a  second (ghostly) scalar mode.

After this choice is made, we are free to add  other derivative operators and write the Lagrangian
density as
\bea
 {\cal L}^{(s)}_{d}&=& \mpl^2 \,b\,\left[ (\delta K_{ij})^2-(\delta K_{ii})^2\right]
 \nonumber\\
&&+   \frac{1}{4}\mpl^2 \Big[d_1(\de_ih_{jk})^2+d_2(\de_ih_{ij})+d_3(\de_ih_{jj})^2 +d_4(\de_ih_{jj}\de_kh_{ik}) 
   \nonumber
   \\
		          && \hskip1.5cm 
		          +c_1(\de_ih_{00}\de_jh_{ij})+c_2(\de_ih_{0i}\de_0h_{jj})+c_3(\de_ih_{00})^2\Big]+
\nonumber
\\
&&+
 \frac{1}{4}a\mpl^3\left[\alpha_1(h_{00}\de_0h_{ii})+\alpha_2(h_{00}\de_ih_{0i})+\alpha_3(h_{ii}\de_jh_{0j})+\alpha_4(h_{ij}\de_ih_{0j})\right] \;.
\eea
Interestingly, also first derivative terms can be added,
however the condition $\alpha_1=2\alpha_2$ in the single derivative sector has to be imposed, in order to avoid the propagation of a second (ghostly) scalar mode.

Collecting these pieces together, 
the new action for the scalars will then be
\begin{eqnarray}\label{S_scal_full}
 S^{(S)}&=\displaystyle\frac{1}{4}\mpl^2\int\dif^4x\,\,&\,\!\!\!\!\!\,\,\,a^2\,\Big\{-6\left(\tau'+aH\psi\right)^2+2\left(2\psi-\tau\right)\nabla^2\tau+4\left(\tau'+aH\psi\right)\nabla^2\left(2v-\sigma'\right) \nonumber\\
  & & + a^2\left[\left(m_0^2+2\epsilon H^2\right)\psi^2-2m_1^2v\nabla^2v-m_2^2\left(\sigma\nabla^4\sigma+2\tau\nabla^2\sigma+3\tau^2\right) \right.\nonumber\\
  & & + \left.m_3^2\left(\nabla^2\sigma+3\tau\right)^2-2m_4^2\psi\left(\nabla\sigma+3\tau\right)\right] \nonumber\\
  & & + b\left(8\tau'\nabla^2v-4\tau'\nabla^2\sigma'-6\tau'^{2}\right) -c_1\nabla^2\psi\left(\nabla^2\sigma+\tau\right) \\
  & & -c_2\nabla^2\psi\left(\nabla^2\sigma+3\tau\right)-c_3\psi\nabla^2\psi -\left(d_1+d_2+d_3+d_4\right)\nabla^2\sigma\nabla^4\sigma\nonumber\\
  & & -2\left(d_1+d_2+3d_3+2d_4\right)\tau\nabla^4\sigma -\left(3d_1+d_2+9d_3+3d_4\right)\tau\nabla^2\tau\nonumber\\
  & & +a \mpl^3\left[\alpha_1\psi(\nabla^2\sigma'+3\tau')+2\alpha_1\psi\nabla^2v+\alpha_3\nabla^2v(\nabla^2\sigma+3\tau)+\alpha_4\nabla^2v(\nabla^2\sigma+\tau)\right]\Big\}\nonumber
\end{eqnarray}
where the parameter $b$ is associated to the combination $( \delta K_{ij})^2-( \delta K_{ii})^2$ expanded
at quadratic order in fluctuations. 
As we said, the fields $v$ and $\psi$ are again auxiliary and their equations of motion can be solved algebraically.
The main point is that
the action resulting from their substitution does not contain any time derivative term $\sigma'$, which means that 
the dangerous `sixth-mode' 
$\sigma$ is not
dynamical and can be integrated away. The action for the only remaining dynamical  scalar
has the following simple structure:
\begin{equation}\label{scal_structure}
 S=\mpl^2\int\dif^4xa^2\left[A_1\tau'^2+A_2\tau\tau'+A_3\tau^2+A_4\sigma^2+A_5\sigma\tau+A_6\sigma\tau'\right] \;,
\end{equation}
where the $A_i$ are functions of all the parameters and the gradient $\nabla^2$ (see Appendix \ref{appB}).
  The field $\sigma$ can then
be integrated out to give (after some integrations by parts)
\begin{equation}
 S=\mpl^2\int\dif^4xa^2\left[B_1\tau'^2+B_2\tau^2\right]\;,
\end{equation}
At this point, one can canonically normalize $\hat\tau=\sqrt{B_1}\tau$ and symbolically expand in $\nabla^2$
(which can be understood in Fourier space as an expansion in the momentum $k$), so that one can read the mass and the
speed of sound of the scalar mode:
\begin{equation}\label{eff_mass_cs}
 S=\int\dif^4x\,a^2\left[\hat\tau'^2 + \hat{c}_s^2\hat\tau\nabla^2\hat\tau+a^2\hat{M}^2\hat\tau^2+\mathcal{O}(\nabla^4)\right] \;.
\end{equation}
The expression of $\hat c_s^2$ and $\hat M^2$ are complicated functions of all the parameters.
It can be checked that in the limit where all the parameters of the modified kinetic terms $b$, $c_i$, $d_i$,
$\delta_i$, $\alpha_i$ vanish, we recover the expressions of the previous section where $c_s$ is given by
Eq.~\eqref{firstcs} and mass is given by 
Eq.~\eqref{firstM}, while higher-order derivative terms correctly drop to zero.
As an example, we write here the effective mass and speed of sound at leading order in slow roll in the case where all
the parameters are zero except for masses and $\alpha_1$:
\begin{eqnarray} 
 \hat{c}_s^2 &=& \frac{\alpha_1 \Lambda  (m_2^2-m_3^2) (\alpha_1 \Lambda -4
   H)}{(m_2^2-m_3^2) \left(3 \alpha_1 \Lambda  (\alpha_1 \Lambda -8 H)+8
   m_0^2\right)+8 m_4^4} \;, \\
 \hat{M}^2   &=& -\frac{m_2^2 (4 H-\alpha_1 \Lambda ) \left(4 H \left(m_2^2-3 m_3^2+3
   m_4^2\right)-\alpha_1 \Lambda  \left(m_2^2-3 m_3^2\right)\right)}{(m_2^2-m_3^2)
   \left(3 \alpha_1 \Lambda  (\alpha_1 \Lambda -8 H)+8 m_0^2\right)+8 m_4^4}\;.
\end{eqnarray}

One can see that `kinetic operators' like the one proportional to $\alpha_1$ can also affect the effective mass.
 A natural question to ask is  whether, by exploiting this fact, 
 effective mass contributions
   can be generated even in the absence of explicit non-derivative terms in the action. This will be the
subject of the next section.

Also after adding derivative contributions, the  curvature perturbation is again not conserved and decays after horizon exit. As previously, this can be seen also from the
trace of the spatial part of the energy-momentum tensor, which, in the simple example we do, now reads
\begin{equation}
 {\rm Tr}\left[\delta T_{ij}\right]= m_2^2\tau +\frac{1}{2}\alpha_1\mpl(a \psi)'+\mathcal{O}(\nabla^2)\;,
\end{equation}
hence it does not vanish at superhorizon scales, due to the contributions proportional to $m_2^2$ and $\alpha_1$. One might use the constraint equation (\ref{eom_vpsi}) to express $\psi'$ in terms of $\tau$, the only 
propagating scalar degree of freedom in the system. It would be interesting to analyze
 how  the curvature perturbation $\zeta$ evolves at superhorizon scales when $\alpha_1$ or other diffeomorphism-breaking
 kinetic terms are included.



\section{Generating a mass without mass: 
   four derivative operators} \label{4derivatives}

We have learned in the previous section that by  breaking spatial  diffeomorphism invariance 
of the action for metric perturbations, by means of 
 mass terms or derivative operators,  
 we
can change  some of the properties of the tensor spectrum with respect to the standard inflationary predictions,
 in particular its tilt $n_T$ and the value of the tensor sound speed $c_T$.

It is natural to ask  
whether it is really necessary to explicitly  break spatial diffeomorphism invariance   to do so. 
The aim of this section is to show that the answer is no, provided that we allow for higher spatial 
derivative operators in the quadratic action for fluctuations.  
 An effective field theory approach to inflation that takes into account of higher derivative operators
 has also  been proposed in \cite{Weinberg:2008hq}.   
Adding such operators, one can avoid
the argument   \cite{Creminelli:2014wna} (based on operators
with at most two space-time derivatives) and find genuine contributions to the tensor sound speed $c_T$,
that cannot be eliminated by disformal transformations.   
This has interesting  implications since the tensor sound speed enters in the amplitude of the tensor  power
 spectrum (see eq \eqref{ptsca}) in a way that enhances the amplitude of 
  ${\cal P}_T$ that scales as $c_T^{-1}$.
 It would be interesting to find explicit models able to avoid the Lyth bound using this fact, but would also need to consider the effect on the scalar modes and hence the observed tensor-to-scalar ratio $r$.


 In particular, we will explore the effect of 4-derivative contributions to the action for fluctuations, organized
 in such a way  as not to break the spatial diffeomorphism invariance, and not to introduce instabilities.  
   The starting point is to consider the quantities
\bea
\partial_0 \partial_l\,h_{ij}&=& \partial_l\,\chi_{ij}'+\partial_l \partial_{(i} s'_{j)}
+\partial_l \partial_i \partial_j \,\sigma'+\delta_{ij}\,\partial_l \tau' \,,
\\
\partial_0 \partial_i\,h_{ij}&=&\nabla^2 s_j'+\partial_j\,\nabla^2\,\sigma'+\partial_j\tau' \,,
\\
\partial_0 \partial_j\,h_{ii}&=&\partial_j\,\nabla^2\,\sigma'+3 \partial_j\tau' \,,
\eea
that we can use to build quadratic operators with four derivatives, that we can potentially add to the action
for metric perturbations
\bea
L_1&=&\left({\partial_l \,\partial_0}\,h_{ij}\right)^2
\,=\,\left( \partial_l\,\chi_{ij}'\right)^2+2
\left( \nabla^2  s'_j\right)^2
- \nabla^2  \,\sigma'\, \nabla^2 \nabla^2 \,\sigma'-
 3\tau'\,\nabla^2 \tau'-2\,\nabla^2  \,\sigma'\,  \nabla^2 \tau'
 \,,
\\
L_2&=&\left( 
\partial_0 \partial_i\,h_{ij}
\right)^2\,=\,
 \left( \nabla^2  s'_j\right)^2
- \nabla^2  \,\sigma'\, \nabla^2 \nabla^2 \,\sigma'-
\tau' \nabla^2 \tau'-2\,\nabla^2  \,\sigma'\,  \nabla^2 \tau'
%
 \,,
\\
L_3&=&
\left( \partial_0 \partial_j\,h_{ii}\right)^2\,=\,
- \nabla^2  \,\sigma'\, \nabla^2 \nabla^2 \,\sigma'-
 9\tau'\,\nabla^2 \tau'-6\,\nabla^2  \,\sigma'\,  \nabla^2 \tau'
  \,,
 \\
L_4&=&\partial_0 \partial_i\,h_{ij} \partial_0 \partial_j\,h_{ii}
\,=\,- \nabla^2  \,\sigma'\, \nabla^2 \nabla^2 \,\sigma'-
 3\tau'\,\nabla^2 \tau'-4\,\nabla^2  \,\sigma'\,  \nabla^2 \tau'
 \,,
 \eea
  where integrations by parts have been  performed.
We would like to build a combination of $L_i$ such that only  contributions
 associated with $\chi_{ij}'\,\nabla^2\,\chi_{ij}'$ and
  $\tau'\nabla^2\tau'$ are non-vanishing, while the vectors and the remaining scalars do not appear.
   If such combination can be found, it is invariant under spatial diffeomorphisms, since $\chi_{ij}$ and $\tau$
   do not transform under this symmetry (see eq \eqref{diffeotr}, noticing that $\tau$ transforms but 
   only under time-reparameterization). 
 The  combination with the desired properties is  
\begin{eqnarray}\label{rhcomb}
 L_{\omega_1} &=& \omega_1(L_1-2L_2-L_3+2L_4) \\
         &=& -\omega_1\chi'_{ij}\nabla^2\chi'_{ij}+2\omega_1\tau'\nabla^2\tau'  
 \,.
\end{eqnarray}
 In analogy to what happens for the two derivatives operators, see the comment after Eq.~(\ref{derct}), this  combination \eqref{rhcomb} corresponds to a particular
combination of the extrinsic curvature perturbation, 
\be
(\de_i\delta K_{jk})^2-(\de_i\delta K)^2-2(\de_i\delta K_{ij})^2-2\de_i\delta K\de_j \delta K_{ij}\,,\ee
 expanded at quadratic order in perturbations.

\bigskip

 Analogously, one can consider four derivative operators that lead
only to combinations involving four spatial derivatives acting on the tensors  $\nabla^2 \chi_{ij}\,\nabla^2\,\chi_{ij}$.
The following Lagrangians arise from all possible contractions of two spatial derivatives and $h_{ij}$ (once integrations by parts
are taken into account):
\begin{eqnarray}
 L_1 &=& (\nabla^2h_{ij})^2 = (\nabla^2\chi_{ij})^2-2s_i\nabla^4s_i+(\nabla^4\sigma)^2+3(\nabla^2\tau)^2+2\nabla^2\tau\nabla^4\sigma  \,,\\
 L_2 &=& (\de_i\de_jh_{ij}) = (\nabla^4\sigma+\nabla^2\tau)^2  \,,\\
 L_3 &=& (\nabla^2h_{ii})^2 = (\nabla^4\sigma+3\nabla^2\tau)^2  \,,\\
 L_4 &=& (\de_k\de_ih_{ij})^2 = -s_i\nabla^4s_i+(\nabla^4\sigma+\nabla^2\tau)^2  \,,\\
 L_5 &=& (\nabla^2h_{kk}\de_i\de_jh_{ij}) = (\nabla^4\sigma+\nabla^2\tau)(\nabla^4\sigma+3\nabla^2\tau)
 \,. 
\end{eqnarray}
There exist combinations of these operators  which allow us to avoid contributions from all vectors and scalars:
\begin{eqnarray}
 L_{\omega_2} &=& \omega_2(L_1+\frac{1}{2}L_2-\frac{1}{2}L_3-2L_4+L_5) =\\
         &=& \omega_2(\nabla^2\chi_{ij})^2 \,,
\end{eqnarray}
hence this combination preserves full four dimensional diffeomorphism invariance. 


By adding the Lagrangians $ L_{\omega_1}$ and $ L_{\omega_2}$ to the quadratic  EH Lagrangian plus the 
two derivative contribution (\ref{derct}) -- that can preserve diffeomorphism invariance if it originates
from a combination of $\delta K_{ij}^2$ and $^{(3)}R$ 
(see the comment after Eq.~(\ref{derct})) --
  one obtains the effective Lagrangian
for tensor modes\footnote{The same operators will also modify the scalar sector. Considering for simplicity only
the Einstein-Hilbert part plus these four-derivative operators, it can be easily seen that the action for the scalar has the same
form of the action for the tensors \eqref{ltquar} and that the arguments that can be developed
 for the scalar sector are very similar to the ones we are carring on for the tensors.
 }
 
\be\label{ltquar}
{\cal L}^{(T)}\,=\,\frac{M_{Pl}^2}{4}\,a^2\,\left[(1+b) ({\chi}'_{ij})^2-
\frac{\omega_1}{a^2 \Lambda^2}\,{\chi}'_{ij}\,\nabla^2
{\chi}'_{ij}+(1+d) \chi_{ij}\,\nabla^2\,\chi_{ij}+
\frac{\omega_2}{a^2 \,\Lambda^2} \, \chi_{ij}\,\nabla^2\,\nabla^2\,\chi_{ij}
\right]
\ee

\smallskip

\noindent
with $\omega_{1,2}$  arbitrary parameters, and $\Lambda$ some  cut-off energy scale, 
 that will depend on the UV completion, and 
that  to  be safe we take larger 
  than  the Hubble scale during inflation.
   Let us emphasize that we constructed the Lagrangians $L_{\omega_1}$ and
     $L_{\omega_2}$ as space diffeomorphism invariant combinations,  with the 
     specific 
      aim to analyze the phenomenological consequences of  higher order derivative operators in the tensor
     sector. These Lagrangians are characterized by a specific choice of parameters among their terms: 
      it would be interesting to 
       investigate whether such combinations can be enforced by some symmetry principle. 
        
    To canonically normalize the tensor field appearing in the Lagragian ${\cal L}^{(T)}$
     of eq. \eqref{ltquar}, 
    we pass for simplicity
to Fourier space, and define the quantity 
\be\label{norchi}
{\chi}_{ij}\,=\,\frac{\sqrt{2}\,\tilde \chi_{ij}}{M_{Pl}\,a\,\sqrt{1+b+\omega_1\,k^2/(a^2 \,\Lambda^2)}}
\,.
\ee
Using this tilde quantity $\tilde {\chi}_{ij} $, 
 the Lagrangian, after 
an integration by parts, 
 acquires a relatively simple form in a quasi-de Sitter universe 
\be
{\cal L}^{(T)}\,=\,\frac{1}{2}\,\left[(\tilde{\chi}_{ij}')^2-
F(k,\eta)
\,\tilde{\chi}_{ij}^2
\right]
\ee
with
\bea
F(k,\,\eta)&=&\frac{1}{\left(1+ b+\frac{\omega_1\,k^2}{a^2\,\Lambda^2}\right)^2}
\,\Big[ -
(1+b)^2\,\left(2-\epsilon\right)\,a^2\,H^2 
+k^2 
\left(1+b\right)\left(1+ d-\left(3-\epsilon\right) \frac{\omega_1 H^2}{\Lambda^2}\right)
\nonumber
\\
&&+\frac{k^4}{a^2\,\Lambda^2}\left( 
\omega_1+ d\,\omega_1+\omega_2+b\,\omega_2
\right)
+\omega_1 \omega_2 \,\frac{k^6}{a^4\,\Lambda^4}
\Big] \,.
\eea

\smallskip

We can now work out some consequences of these results:

\smallskip
\noindent
$\bullet$
   By making the choice $b=-1$,   
  the quadratic terms containing two time derivatives cancel from the action (\ref{ltquar}), and the dynamics
  is driven by the four derivative operator proportional to $\omega_1$. In a certain sense, the situation can be seen
  as analogous 
    to what happens in ghost inflation \cite{ArkaniHamed:2003uz}, where 
     the leading terms in the gradients of the ghost field vanish, and the  next-to-leading contributions in gradients
      become dominant.

 The expression for the function  $F$ above simplifies considerably:
 \bea
 F(k,\,\eta)&=&\frac{\omega_2}{\omega_1}\,k^2+\frac{(1+d)\,\Lambda^2}{\omega_1}\,a^2
 \,,
\\
&=&\frac{\omega_2}{\omega_1}\,k^2-2 \,H^2\,a^2+\frac{(1+d)\,\Lambda^2+2\,H^2\,\omega_1}{\omega_1}\,a^2
  \,. 
 \label{simf}
 \eea
The first term in the right hand side
of (\ref{simf}) can be recognized as the usual
 first contribution to the dispersion relation associated with $\tilde{\chi}_{ij}$, characterized by  an effective  
 sound speed  $c_T^2\,=\omega_1/\omega_2$. 
  The second piece is the effective `mass term' that usually arises in a quasi-de Sitter universe. 
 Then, 
 we have the third contribution, that mimics exactly a  mass term  with 
 \be
 m_{\tilde \chi}^2=\frac{(1+d)\,\Lambda^2+2\,H^2\,\omega_1}{\omega_1}\,.
 \ee
 Interestingly this effective mass   arises only
 from the higher derivative terms, with no need to break diffeomorphism invariance!
 In this sense,  4-derivative contributions can be interpreted as being able  to generate mass without an explicit  mass
  parameter. On the other hand, notice that in this case
 the relation between the canonically normalized  tensor field $\tilde \chi_{ij}$ and original one $\chi_{ij}$
  scales as the inverse
 of the momentum: $\chi_{ij}\propto \tilde{\chi}_{ij}/k$: see eq. (\ref{norchi}). This typically implies -- by the arguments
 outlined around eq \eqref{cocond} -- a low cut-off scale when focussing
  at large scales; on the other hand,  this crucially depends  on the 
 tensor interactions during inflation, that might conspire in such a way to raise the cut-off. This is an interesting question that we intend to pursue in the future.

\noindent
$\bullet$
Let us now consider the more general situation with $b\neq-1$, focusing on the large and small scale limits
for the function $F$:
\begin{eqnarray}
 F(k,\,\eta) & \stackrel{k\to0}{\sim}       & (-2+\epsilon)a^2H^2
 +\mathcal{O}(k^2)  \,,\\
 F(k,\,\eta) & \stackrel{k\to+\infty}{\sim} & \frac{\omega_2}{\omega_1}k^2+\frac{a^2\Lambda^2}{\omega_1^2}\left[(1+d)\omega_1-(1+b)\omega_2\right]+\mathcal{O}(k^{-2})
 \,.  
\end{eqnarray}
No major differences with respect to  the standard case arise, apart from the
presence of a non-trivial sound speed $c_T$: the system can be quantized selecting a Bunch-Davies vacuum 
at very small scales,  while at large scales the tensors behave as in a standard quasi-de Sitter universe, 
with no mass. 


\bigskip

This preliminary analysis
 of the role 
of operators with higher spatial derivatives shows their    
 possible relevance for characterizing tensor modes, and can find some motivation for example (but not only) in the context
 of Horava-Lifshitz cosmology (see \cite{Mukohyama:2010xz}  for a review). It shows that in this set-up a non-unity
 tensor sound speed $c_T$ can be generated,  and that it cannot in general be set to one by a set of transformations
 of the metric \cite{Creminelli:2014wna}.  

 \section{Conclusions}

  By implementing
 an effective field theory approach to single clock inflation, 
 we have examined interesting properties of the spectrum of inflationary tensor fluctuations, that arise 
 when breaking some of the symmetries or requirements  usually imposed on  the dynamics of  inflationary perturbations.
 %

    In the first part of the paper we considered  the possibility  that, besides time-reparameterization, spatial diffeomorphisms are also broken in the quadratic Lagrangian controlling
      fluctuations during inflation. We do so considering  quadratic  operators 
       that break spatial diffeomorphisms,
        maintaining
      spatial isotropy and homogeneity, 
         that 
      contain at most two space-time derivatives.       Such operators can be motivated by a modification of gravity during the inflationary era, or by 
 some particular behavior        of the fields 
      that drive inflation.    
      We identified the single operator that contributes at leading order to the tensor spectral tilt $n_T$, and that can change its sign leading to a positive $n_T$ without necessarily violating the null energy 
condition.       We have then shown that this operator has important consequences in the scalar sector. It generically leads 
to superhorizon non-conservation   of the curvature perturbation $\zeta$ on uniform energy density slices, even in single clock inflation -- since $\zeta$ acquires an effective mass -- although additional allowed operators can render
the  mass of  $\zeta$ (and its
  non-conservation after horizon exit)  arbitrary small.

      In the second part of the paper, we returned to the case of spatial diffeomorphism invariant Lagrangians, 
  including quadratic 
      operators with more than two spatial derivatives (but no more than two time derivatives)  acting on the tensor perturbations. We showed that also in this
       case, by a judicious choice of the operators,  one can obtain properties for the fluctuations   that are very similar to the ones
        of a diffeomorphism breaking set-up.
         In particular,   a non-trivial tensor sound speed can be
          generated, and the formula for $n_T$ receives  new  contributions that depend  on
             the coefficients of these higher derivative operators. We also discussed
             a special case in which  
                  such operators can mimic the effect 
          of a mass term in the tensor sector. 
     
     
     The power of our approach is the use of effective field theory of inflation \cite{Cheung:2007st}, that relies on symmetry principles only, 
     and encompasses various scenarios  in  a model independent way. 
      In a companion work \cite{preparation}, using again an effective field theory approach,  we will examine model independent  consequences of breaking isotropy and homogeneity in the Lagrangian for cosmological fluctuations. 
      
      In this work, for simplicity we focussed on a quadratic action for fluctuations since when we break symmetries such as spatial diffeomorphism invariance, operators cubic or higher in fluctuations exist in large number. 
     It  would be interesting to extend our analysis to higher order in perturbations, to study the consequences for non-linearity and non-Gaussianity of the primordial metric perturbations from inflation.

\acknowledgments
It is a pleasure to thank 
Hassan Firouzjahi, Eichiiro Komatsu,  Kazuya Koyama,
 Azadeh Maleknejad,  
 and Ivonne Zavala for useful discussions. 
 GT is supported by an STFC Advanced Fellowship ST/H005498/1. DW is supported by STFC grants ST/K00090X/1
 and ST/L005573/1.  
\appendix

\section{Combinations of $h$ and derivatives}\label{appA}

Combinations up to second order in $h$ and up to two derivatives, avoiding time derivatives on $N$ or $N^i$ (some integrations
by parts have already been performed).
\begin{eqnarray}
 h_{00}\de_0h_{ii}      &=& \psi(\nabla^2\sigma'+3\tau') \label{alpha1}
 \eea
 \bea
 h_{00}\de_ih_{0i}      &=& \psi\nabla^2v \label{alpha2}
 \eea
 \bea
 h_{ii}\de_jh_{0j}      &=& \nabla^2v(\nabla^2\sigma+3\tau) \label{alpha3}
\eea
\bea
 h_{ij}\de_ih_{0j}      &=& \nabla^2v(\nabla^2\sigma+\tau) -u_i\nabla^2s_i \label{alpha4}
       \end{eqnarray}
\begin{eqnarray}       
 (\de_ih_{00})^2        &=& -\psi\nabla^2\psi \eea
 \bea
 (\de_0h_{ii})^2        &=& (\nabla^2\sigma'+3\tau')^2 \eea \bea
 (\de_ih_{0i})^2        &=& (\nabla^2v)^2  \eea \bea
 \de_ih_{0i}\de_0h_{jj} &=&\nabla^2v(\nabla\sigma'+3\tau') \eea \bea
 (\de_ih_{jj})^2        &=& -(\nabla^2\sigma+3\tau)\nabla^2(\nabla^2\sigma+3\tau)  \eea \bea
 (\de_ih_{ij})^2        &=& -(\nabla^2\sigma+\tau)\nabla^2(\nabla^2\sigma+\tau) +(\nabla^2s_j)^2 \label{d2}  \eea \bea
 \de_ih_{jj}\de_kh_{ik} &=& -(\nabla^2\sigma+3\tau)\nabla^2(\nabla^2\sigma+\tau)  \eea \bea
 \de_ih_{00}\de_ih_{jj} &=& -\nabla^2\psi(\nabla^2\sigma+3\tau)  \eea \bea
 \de_ih_{00}\de_jh_{ij} &=& -\nabla^2\psi(\nabla^2\sigma+\tau)  \eea \bea
 \de_jh_{0i}\de_0h_{ij} &=& \nabla^2v(\nabla^2\sigma'+\tau')-u_i\nabla^2s_i' \label{b3}  \eea \bea
 (\de_0h_{ij})^2        &=& (\chi'_{ij})^2+(\nabla^2\sigma')^2+2\tau'\nabla^2\sigma'+3{\tau'}^2-2s_j'\nabla^2s_j' \label{b1} \eea \bea
 (\de_ih_{0j})^2        &=& (\de_iu_j)^2+(\nabla^2v)^2 \label{b2} \eea \bea
 (\de_ih_{jk})^2        &=& (\de_i\chi_{jk})^2+(\de_i\de_j\de_k\sigma)^2-2\nabla^2\sigma\nabla^2\tau-3\tau\nabla^2\tau+2(\nabla^2s_i)^2 \label{d1}
\end{eqnarray}

\section{Speed of sound and mass}\label{appB}

Coefficients $A_i$ for the scalar action \eqref{scal_structure}

\begin{eqnarray}
 A_1 &=& -\frac{\mpl^2(1+b)}{2(\alpha_1\Lambda-4H)^2}\left[-8(1+b)\left(c_3k^2+(m_0^2+2\epsilon H^2)\right)+48baH-3a^2\alpha_1\Lambda(\alpha_1\Lambda-8H)\right] \\
 A_2 &=& \frac{a\mpl^2(1+b)}{(\alpha_1\Lambda-4H)^2}\bigg\{\Big[(3c_2+c_1-4)(\alpha_1\Lambda-4H)+c_3(3\alpha_3+\alpha_4)\Lambda\Big]k^2+\nonumber\\
     & &  \left.+\left[\left(m_0^2+2\epsilon H^2-\frac{6bH^2}{1+b}\right)(3\alpha_3+\alpha_4)\Lambda-6m_4^2(\alpha_1\Lambda-4H)\right]\right\}+ \\
     & &  +\frac{a^3\mpl^2\alpha_1(3\alpha_3+\alpha_4)(\alpha_1\Lambda-8H)\Lambda^2}{8(\alpha_1\Lambda-4H)}\nonumber\\
 A_3 &=& \frac{a^2\mpl^2k^2}{(\alpha_1\Lambda-4H)^2}\Big[4(2+3d_1+d_2+9d_3+d_4)(\alpha_1\Lambda-4H)^2\nonumber\\
     & & +(3\alpha_3+\alpha_4)(2(3c_2+c_1-4)(\alpha_1\Lambda-4H)+c_3(3\alpha_3+\alpha_4))\Lambda\Big]+\nonumber\\
     & & +\frac{a^4\mpl^2}{(\alpha_1\Lambda-4H)^2}\left[6H(m_2^2-3m_3^2)(\alpha_1-2H)+3(3\alpha+\alpha_4)m_4^2H\Lambda\right]+ \\
     & & -\frac{a^4\mpl^2\Lambda^2}{16(\alpha_1\Lambda-4H)^2}\left[12\alpha_1^2(m_2^2-m_3^2)+(3\alpha_3+\alpha_4)(12\alpha_1m_4^2-(3\alpha_3+\alpha_4)(m_0^2+2\epsilon H^2-6H^2))\right]\nonumber\\
 A_4 &=& \frac{a^2\mpl^2k^6}{16(\alpha_1\Lambda-4H)^2}\Big[4(d_1+d_2+d_3+d_4)(\alpha_1\Lambda-4H)^2+2(\alpha_3+\alpha_4)(c_1+c_2)(\alpha_1\Lambda-4H)\Lambda+c_3(\alpha_3+\alpha_4)^2\Lambda^2\Big] \nonumber\\
     & & -\frac{a^4\mpl^2k^4}{16(\alpha_1\Lambda-4H)^2}\Big[4(m_2^2-m_3^2)(\alpha_1\Lambda-4H)^2+4m_4^2(\alpha_1\Lambda-4H)(\alpha_3+\alpha_4)\Lambda+\\
     & & +(m_0^2+2\epsilon H^2-6H^2)(\alpha_3+\alpha_4)^2\Lambda^2\Big]\nonumber\\
 A_5 &=& -\frac{a^2\mpl^2k^4}{8(\alpha_1\Lambda-4H)^2}\Big[4(d_1+d_2+5d_4)(\alpha_1\Lambda-4H)^2-c_3(\alpha_3+\alpha_4)(3\alpha_3+2\alpha_4)\Lambda^2\nonumber\\
     & & -2\alpha_3(\alpha_1\Lambda-4H)(3c_2+2c_1-2)\Lambda-2\alpha_4(\alpha_1\Lambda-4H)(2c_2+c_1-1)\Lambda\Big]\nonumber\\
     & & +\frac{a^4\mpl^2k^2}{16(\alpha_1\Lambda-4H)^2}\Big[4(m_2^2-3m_3^2)(\alpha_1\Lambda-4H)^2 \\
     & & -(m_0^2+2\epsilon H^2-6H^2)(\alpha_3+\alpha_4)(3\alpha_3+2\alpha_4)\Lambda^2+4m_4^2(\alpha_1\Lambda-4H)(3\alpha_3+2\alpha_4)\Lambda\Big]\nonumber\\
 A_6 &=& -\frac{a^2\mpl^2k^4}{8(\alpha_1\Lambda-4H)}\Big[(c_1+c_2)(\alpha_1\Lambda-4H)-c_3(\alpha_3+\alpha_4)\Lambda\Big]+\nonumber\\
     & & +\frac{a^3\mpl^2k^2}{8(\alpha_1\Lambda-4H)}\big[16m_4^2(1+b)(\alpha_1\Lambda-4H)-8(1+b)(m_0^2+2\epsilon H^2)(\alpha_3+\alpha_4)\Lambda+\nonumber\\
     & & +3(16bH^2-\alpha_1\Lambda(\alpha_1\Lambda-8H))(\alpha_3+\alpha_4)\Lambda\big]
\end{eqnarray}




\bibliography{bibliography}

\end{document}